\documentclass[superscriptaddress,aps,prd,reprint]{revtex4-2}
 
\usepackage[T1]{fontenc}
 
\usepackage{graphicx}
\usepackage{dcolumn}
\usepackage{bm}
 
\usepackage{amsfonts}
\usepackage{amssymb}
\usepackage{amsmath}
\usepackage{mathtools}
\usepackage{mathrsfs}
\usepackage{physics}
 
\usepackage{slashed}
\usepackage{cancel}
\usepackage{color}
\usepackage{enumitem}

\usepackage{flushend}
\usepackage[hyperfootnotes=false]{hyperref}
\hypersetup{colorlinks=true, linkcolor=magenta, filecolor=cyan, urlcolor=blue}
\usepackage{cleveref}

\def\bi{\begin{itemize}}
\def\ei{\end{itemize}}
\def\be{\begin{equation}}
\def\ee{\end{equation}}
\newcommand{\bea}{\begin{eqnarray}}
\newcommand{\eea}{\end{eqnarray}}
\AtBeginDocument{%
}

\begin{document}

\title{Symbiotic Magnetogenesis during Radiation Domination}

\author{Stephon Alexander}
\email{stephon\_alexander@brown.edu}

\author{Ariel Marxena Baksh}
\email{ariel\_baksh@brown.edu}

\author{Lawrence Edmond IV}
\email{lawrence\_edmond@brown.edu}

\author{Wenrong Sun}
\email{wenrong\_sun@brown.edu}

\affiliation{Department of Physics, Brown University, Providence, RI 02912, USA}
\affiliation{Brown Center for Theoretical Physics and Innovation (BCTPI), Providence, RI 02912, USA}

\date{\today}

\begin{abstract}
\noindent
We present a late-time magnetogenesis mechanism in which a coupled axion-dilaton system sources a dark $U(1)$ gauge field. The dilaton’s exponential coupling drives tachyonic amplification by reshaping the instability band, while the axion controls the helicity structure of the field. Together, they amplify both gauge helicities and produce a moderately chiral dark magnetic field without fine-tuning in either scalar sector. The field is generated in the dark sector, thus the mechanism avoids plasma-conductivity suppression in the visible sector, while the model remains robust across a broad range of scalar-masses and couplings. Numerical evolution from $z=10^5$ to matter-radiation equality, combined with a parameter search over the axion mass, dilaton initial conditions, and dilaton coupling, shows that astrophysically relevant amplification persists across fuzzy-dark-matter and ultralight-axion regimes. A benchmark case yields $B\approx0.9 ~\mathrm{nG}$ on $\lambda_0\sim1 \mathrm{Mpc}$, with kinetic mixing transferring the field to the visible sector over a broad range of mixing parameters.

\end{abstract}

\maketitle

%%%%%%%%%%%%%%%%%%%%%%%%%%%%%%%%%%%%%%%
\section{Introduction}
%%%%%%%%%%%%%%%%%%%%%%%%%%%%%%%%%%%%%%%
\par
Observations of $\mathrm{TeV}$ blazar spectra and gamma-ray cascades suggest a lower bound of $B_{\lambda} \gtrsim 10^{-17} \, \mathrm{G}$ on the amplitude of intergalactic magnetic fields, while measurements of the Cosmic Microwave Background (CMB) further constrain these amplitudes to $B_{\lambda} \lesssim 10^{-9} \, \mathrm{G}$, for length scales $\lambda_c \lesssim 1.0 \,\mathrm{Mpc}$~\cite{Widrow:2002ud, Wang:2020vyu, Burmeister:2025lgo, 2011A&A...529A.144T, Barrow:1997mj, Biteau:2022dtt, Kahniashvili:2015msa, Subramanian:2003sh, Uchida:2024iog, HESS:2023zwb}. The detection of these fields in voids without apparent astrophysical sources capable of sustaining a dynamo strongly suggests they may be of primordial origin. Their generation and subsequent evolution remains an open question for cosmologists and phenomenologists~\cite{Durrer:2013pga}. Many theoretical models have been proposed in which the electromagnetic sector is coupled to scalar or pseudoscalar fields to break conformal invariance in an expanding universe. Models for magnetogenesis during inflation typically include couplings in which the gauge kinetic term is modulated by a function of a scalar field~\cite{Ratra:1991bn,Turner:1987bw,Bamba:2003av,
Martin:2007ue,Demozzi:2009fu,Ferreira:2014hma}. Axion-like pseudoscalar coupling models include odd-parity interactions that can drive helical amplification of the gauge fields~\cite{Caprini:2014mja,Fujita:2015saa,Adshead:2016iae,
Figueroa:2023oxc,Figueroa:2024rkr, Brahma:2026fre}, and arise naturally in multi-axion effective theories with kinetic mixing and alignment~\cite{PhysRevD.91.023520, alexander24axiverse, Mehta:2020kwu}. Late-time post-inflationary mechanisms have also been explored, including resonant axion-$U(1)$ conversion~\cite{Choi:2018dqr, Kitajima:2023pby} and ultralight dark matter seeding of galactic magnetic fields~\cite{Berger:2025pfv, Brandenberger:2025gks, Kushwaha:2026aab, PhysRevD.16.1791, Alexander:2026fam}.

\par
Axion-driven magnetogenesis has received particular attention because it can avoid difficulties that may arise from strong couplings, while still producing maximally helical magnetic fields on cosmological scales~\cite{Caprini:2014mja, Fujita:2015saa}. In this framework, a coherently oscillating pseudoscalar field induces a tachyonic instability for one helicity mode, leading to the exponential amplification of the gauge field within a finite momentum band~\cite{garretson92, Caprini:2014mja}. Lattice simulations confirm this picture by demonstrating significant gauge field production and resolving the helical structure of the resulting fields~\cite{Adshead:2016iae, Figueroa:2023oxc, Figueroa:2024rkr}. However, the mechanism faces important challenges: strong backreaction on the axion background can hinder sustained amplification~\cite{Figueroa:2023oxc, Brahma:2026fre}. Additionally, late-time scenarios are highly susceptible to the conductivity of the photon-baryon plasma during radiation domination~\cite{PhysRevD.56.5254}, placing any visible gauge sector in a strongly overdamped regime, washing out any amplification the axion would otherwise produce. Finally, achieving a sufficiently large magnetic field today with pseudoscalar-coupling to the Standard Model photon at late times generally requires large initial misalignment angles or fine-tuned initial conditions~\cite{Fujita:2015saa}. 

\par
In models with dilaton-like couplings, the coupling function enters the gauge kinetic term multiplicatively, so that the gauge field amplitude inherits an exponential factor that is present at all times. Because the dilaton field value enters the amplitude of the resulting magnetic field today exponentially, even moderate values for the dilaton field can drive substantial amplification that a linearly coupled model could not achieve alone. Furthermore, the dilaton field can contribute an oscillatory friction term to the gauge equations of motion. This additional drag on the gauge fields both reshapes the instability band and allows for the growth of both helicity modes simultaneously. However, dilaton-like models face problems with strong couplings, where the functional coupling $I$ becomes non-perturbatively large at early times if it is small today, or vice versa~\cite{Demozzi:2009fu, Bamba:2003av}. In late-time dilaton-only mechanisms, the dilatonic contribution to the drag on the gauge fields must now compete against Hubble friction on the fields, in addition to the aforementioned challenges with the conductivity of the plasma if the gauge sector is coupled to the Standard Model~\cite{Brandenberger:2025gks}. 

\par
Late-time dilaton-driven magnetogenesis has been studied in Ratra-like models during radiation domination~\cite{Brandenberger:2025gks, Kamali:2026tgq}, finding an extended parameter space that is capable of seeding magnetic fields within current observational constraints, while placing weak constraints on the sourcing dark matter mass. A qualitatively different class of models avoids these limitations by introducing kinetic mixing between a dark photon and the Standard Model photon~\cite{Choi:2018dqr}, where the dark sector develops a large gauge field amplitude through resonant axion-dark photon conversion, which is then transferred to the visible sector electromagnetically. While these mechanisms are compelling, they each rely on a single scalar sector, and thus cannot simultaneously provide both the tachyonic helicity pump and the exponential kinetic enhancement. Combining both couplings into a single model opens the possibility of a richer instability structure that is more robust against the limitations of either mechanism in isolation. 

\par
In this work, we propose a late-time magnetogenesis mechanism based on a coupled axion-dilaton system acting on a dark $U\left(1\right)$ gauge sector, building on the late-time dilaton framework of~\cite{Brandenberger:2025gks}. This model includes both a pseudoscalar coupling and a multiplicative dilaton kinetic coupling. Our numerical analysis demonstrates three key features that distinguish this model from single-field mechanisms. First, the exponential dilaton coupling $I\left(\chi\right)=\exp\left(2\beta \chi/M\right)$ amplifies the magnetic field today $B_{\pm,0}^2 \propto I_*$, where $I_*=I\left(\chi_*\right)$ is the value of the kinetic coupling at generation, $M$ is the dilaton energy scale, and $\beta$ is a dimensionless coupling constant. This allows moderate, cosmologically viable initial conditions $\chi_0 \leq M_\mathrm{Pl}$ to drive substantial amplification without fine-tuning the axion sector. Second, the combined instability band supports the tachyonic growth of \emph{both} gauge helicity modes simultaneously, yielding a moderately chiral field rather than a maximally helical field configuration. Third, placing the gauge sector in a dark $U\left(1\right)_D$ gauge group removes the conductivity obstruction entirely, allowing for the instability to proceed unimpeded during late times. The dark magnetic field is subsequently transferred to the visible sector through kinetic mixing between the dark photon and the Standard Model photon~\cite{Holdom:1986eq}. This induces a visible-sector magnetic field on the same coherence length scale as the dark field. A parameter survey over the dilaton initial conditions, the dilaton coupling $\beta$, and the axion mass (Figs.~\ref{fig:axi_dilaton_energy_density}--\ref{fig:2D_chi0_theta0_parameter_search}) confirms that the amplification sufficient to seed the observed intergalactic magnetic fields is an extended feature of the parameter space. 

\par
The remainder of this work is organized as follows. In Sec.~\ref{sec:the_model} we present the effective Lagrangian and derive the equations of motion for the scalar and gauge sectors, including the rescaled gauge equation and the analytic structure of the tachyonic instability band. In Sec.~\ref{sec:dark_magnetic_fields} we describe the numerical evolution scheme and initial conditions. We also present the results of our parameter search and discuss the dark magnetic field amplitude today, energy density constraints, and the helicity structure. In Sec.~\ref{sec:axion_mass} we present a parameter scan over the axion mass and misalignment angle, and discuss the observational regime in which the mechanism is most distinctive. We conclude in Sec.~\ref{sec:conclusion} with a discussion of observational signatures and directions for future work. 

%%%%%%%%%%%%%%%%%%%%%%%%%%%%%%%%%%%%%%%
\section{The Model}
\label{sec:the_model}
%%%%%%%%%%%%%%%%%%%%%%%%%%%%%%%%%%%%%%%
We begin with the model, 
\begin{align}
\label{Lagrangian}
    \mathcal{L} = &-\frac{1}{2}\left(\partial \phi\right)^2 - V\left(\phi\right) -\frac{1}{2}\left(\partial \chi\right)^2 - U\left(\chi\right) \nonumber \\
    &- \frac{1}{4} I\left(\chi\right)F_{\mu \nu} F^{\mu \nu} -  \frac{\lambda}{4} I\left(\chi\right) \theta_\phi F_{\mu \nu} \tilde{F}^{\mu \nu}
\end{align}
where $\theta_\phi \equiv \phi/f_\phi$ is the scalar misalignment angle with action $S = \int d^4x \sqrt{-g} \mathcal{L}$. The potential for the dilaton is quadratic~\cite{banerjee25}, and for the axion we use a cosine potential~\cite{PhysRevD.16.1791, garretson92}. For the multiplicative coupling, we choose the dimensionless functional $I\left(\chi\right)=\exp\left(2 \beta \chi /M\right)$, where the dimensionless coupling constant $\beta$ can take on both positive and negative values~\cite{Ratra:1991bn, Bamba:2003av}. In the phenomenologically motivated case during radiation domination where this $U(1)$
is identified with a dark photon of mass $m_{\gamma'}$, the mass term enters as an additive $m_{\gamma'}^2 A_\pm$ contribution to the gauge equations of motion, leading to a $+m_{\gamma'}^2$ shift in the dispersion relation. In the regime $m_{\gamma'} \ll m_\phi$, this is negligible compared to the tachyonic growth rate during the instability, and the longitudinal mode decouples in the limit $m_{\gamma'}/k \to 0$~\cite{Graham:2015rva}. Both conditions are satisfied for all parameter values considered in this work. 

\par
We work in a spatially flat Friedmann-Lemaitre-Robertson-Walker (FLRW) cosmology with metric $g^{\mu \nu} = \left(-1, a^2, a^2, a^2\right)$. The equations of motion for the scalar fields are, 
\begin{align}
    \theta_\phi'' &+ 3\frac{H}{m_\phi} \theta_\phi'+ \frac{1}{f_\phi^2 m_\phi^2} \frac{\partial V\left(f_\phi \theta_\phi \right)}{\partial \theta_\phi} = - \frac{\lambda}{4 f_\phi^2 m_\phi^2} I \left(\chi\right) F_{\mu \nu} \tilde{F}^{\mu \nu} \\[5pt]
    \chi'' &+ 3 \frac{H}{m_\phi} \chi' + \frac{1}{m_\phi^2} \frac{\partial U}{\partial \chi} = -\frac{1}{m_\phi^2} \frac{\partial I}{\partial \chi} \left(\frac{1}{4} F^2 + \frac{\lambda}{4} \theta_\phi F \tilde{F}\right) 
\end{align}
where the primes denote differentiation with respect to the dimensionless time parameter $x_\phi = m_\phi t$. Backreaction for both cases is negligibly small compared to the corresponding scalar field potential term consistent with recent analyses of resonance and backreaction in ultralight-dark-matter magnetogenesis~\cite{Brahma:2026fre, Choi:2018dqr}. The detailed analysis is displayed in Appendix~\ref{app:backreaction}. The equations of motion for the gauge field in dimensionless time are, 
\begin{equation}
    A''_\pm + \Gamma\left(x_\phi\right) A'_\pm + \omega^2_{\pm,k}\left(x_\phi\right) A_\pm = 0 
\end{equation}
where the friction term $\Gamma \equiv H/m_\phi + I'/I$ and $\omega^2_{\pm,k}$ is the time-dependent frequency of the gauge fields. We rescale using, 
\begin{equation}
    u_\pm = e^{\frac{1}{2}\int^{x_\phi} \Gamma\left(\tilde{x}\right)d\tilde{x}}A_\pm,
\end{equation}
so that the gauge equations of motion can be written compactly as, 
\begin{equation}
\label{eq:gauge_eom_rescaled}
    u''_\pm + \Omega_{\pm,k}^2 u_\pm = 0 
\end{equation}
with dispersion relation, 
\begin{equation}
\label{eq:Floquet_frequency}
    \Omega_\pm^2\left(x\right) \equiv \left(\frac{k}{a m_\phi}\right)^2 \mp \lambda \left(\frac{k}{a m_\phi}\right) \left(\theta_\phi' + \theta_\phi \frac{I'}{I}\right) -\frac{1}{2}\Gamma' - \frac{1}{4}\Gamma^2.
\end{equation}

\par
Tachyonic instability occurs when $\Omega_{\pm,k}^2 < 0$, driving the amplification of the corresponding gauge mode. To begin studying the instability structure analytically, we work in the coherently oscillating regime where both scalar fields oscillate as decaying sinusoids, 
\begin{equation}
\label{eq:Oscillations}
    \theta\left(x_\phi\right) = \Theta\left(x_\phi\right) \cos\left(x_\phi\right), \,\,\,\, \chi\left(x_\phi\right) = \Xi\left(x_\phi\right) \cos\left(r x_\phi\right),
\end{equation} 
where $r\equiv m_\chi/m_\phi$, and the slowly-varying envelopes decay as $\Theta, \, \Xi \propto a^{-3/2}$. During radiation domination, $H/m_\phi, \, H/m_\chi \ll 1$, so the drag on the gauge fields is dominated by the friction from the dilaton, which is well approximated by, 
\begin{equation}
\label{eq:GammaApprox}
    \Gamma \approx \frac{I'}{I} \approx - \frac{2\beta r}{M} \Xi\left(x_\phi\right) \sin\left(r x_\phi\right).
\end{equation}
We define the helicity and dilatonic pumps as $S\left(x_\phi\right)=\theta' + \theta \Gamma$ and $C\left(x_\phi\right)= \frac{1}{2}\Gamma' + \frac{1}{4}\Gamma^2$, respectively. In this form, the gauge dispersion relation can be written as a quadratic in $\kappa$, 
\begin{equation}
    \Omega_{\pm,k}^2 = \kappa^2 \mp \lambda S\left(x_\phi\right) \kappa - C\left(x_\phi\right),
\end{equation}
where $\kappa \equiv k/\left(a m_\phi\right)$. The boundaries of the instability band follow from setting $\Omega_{\pm,k}^2 = 0$ and solving the resulting quadratic equation to find, 
\begin{equation}
\label{eq:dispersion_relation_quadratic_solution}
    \kappa_\pm \left(x_\phi \right) = \frac{1}{2}\left[\pm \lambda S \pm \sqrt{\lambda^2 S^2 + 4 C}\right].
\end{equation}
The oscillations of the dilatonic pump $C\left(x_\phi\right)$ divide the boundary of the instability band into three specific regimes: $C = 0$, $C > 0$, and $C < 0$.

\par
In the first regime, when the dilatonic pump is turned off (i.e. $C=0$ and $\Gamma = \Gamma' =0$), $S\simeq \theta'$ and $\Omega^2 = \kappa^2 \mp \lambda \theta' \kappa$. The instability band therefore spans $0<\kappa < \abs{\lambda \theta'}$, recovering the standard axion-driven mechanism for chiral amplification via tachyonic instability. In the second regime, $C>0$, so $\Omega^2\left(\kappa=0\right)=-C <0$, meaning the instability band includes $\kappa=0$ and lies within $0 < \kappa < \kappa_+$. The instability band requires $\lambda^2 S^2 +4C \geq 0$, which ensures the discriminant of Eq.~\eqref{eq:dispersion_relation_quadratic_solution} remains real. The helicity pump driven by the axion is out of phase compared to the dilatonic pump. Away from the axion nodes, the discriminant yields the condition, 
\begin{equation}
    \lambda^2 \Theta^2 \sin^2\left(x_\phi\right) \gtrsim \frac{\beta r}{M} \Xi \left(1+2 \lambda^2 \Theta^2\right)
\end{equation}
meaning that in the $0<\kappa <\kappa_+$ band, the discriminant is generically positive and dominated by $\lambda^2 \Theta^2 \sin^2\left(x_\phi\right)$. Near the tachyonic burst, when $\abs{\sin x_\phi} \sim 1$, this reduces to, 
\begin{equation}
    \lambda^2 \Theta^2 \gtrsim \frac{\beta r}{M} \Xi
\end{equation}
which is satisfied given the Planck-suppression of the dilaton contribution. In the third regime, $C < 0$, the instability structure shifts qualitatively. 

\par
When $C<0$, $\Omega^2\left(\kappa=0\right)=-C >0$, so long-wavelength modes are now stable, ultimately pushing the instability band to larger values of $\kappa$ (i.e. smaller physical length scales). If the discriminant remains positive, the resulting tachyonic band lies within $\kappa_- < \kappa < \kappa_+$. The width of the band is $\kappa_+ - \kappa_- = \sqrt{\lambda^2 S^2 + 4C}$ provided $\lambda^2 S^2 + 4C \geq 0$. The existence of the instability band therefore requires $\lambda^2 S^2 \geq 4 \abs{C}$. Near the nodes of the axion cycle, when $\sin x_\phi \approx 0$ and $\cos x_\phi \approx \pm1$, the leading axion velocity pump $\theta'$ vanishes and the helicity pump is driven entirely by the Planck-suppressed dilaton-modulated term, 
\begin{equation}
    S \approx - \frac{2 \beta r}{M} \Theta \Xi \sin\left(r x_\phi\right). 
\end{equation}
Consequently, the finite-$\kappa$ band does not vanish exactly at the axion nodes, but is rather modulated by the dilaton, provided the condition $\lambda^2 S^2 \geq 4 \abs{C}$ remains satisfied. In practice, the instability is strongest when the axion velocity dominates, and is significantly weakened near the nodes where the helicity pump is controlled by the dilaton-induced correction $\theta \, \Gamma$. 

\par
To make the frequency mixing structure explicit, we substitute the coherent oscillation forms directly into Eq.~\eqref{eq:Floquet_frequency} and expand to leading order in the scalar field amplitudes, yielding Eq.~\eqref{eq:axi_dilaton_dispersion_approx}. The first term corresponds to the usual dispersion relation, the second term encodes the helicity-dependent axion-driven instability, and the remaining terms arise from induced drag from the dilaton. In particular, the dilaton introduces additional oscillatory structure through frequency mixing, which modifies and reshapes the instability band.
\begin{widetext}
\begin{equation}
\label{eq:axi_dilaton_dispersion_approx}
    \Omega_{\pm,k}^2 \simeq \kappa^2  \pm \lambda \kappa \Theta \sin x \pm \lambda \kappa \frac{\beta r}{M}\Theta \Xi \left[\sin\left(\left(r+1\right)x\right) + \sin\left(\left(r - 1\right)x\right)\right] + \frac{\beta r^2}{M} \Xi \cos\left(rx\right).
\end{equation}
\end{widetext}
In the tachyonic regime, the gauge-field modes grow exponentially rather than oscillating. When the dilaton contribution is small, this reduces to the standard axion-driven instability. However, when the dilaton is significant, it can either enhance or suppress the instability depending on its phase and amplitude, and can shift the instability to different momentum scales. This approximation holds across the fuzzy dark matter and ultralight axion mass range, where $H/m_\phi \ll 1$ is generically satisfied during radiation domination, ensuring the dilaton friction dominates over Hubble drag throughout the instability. 

\par
Inverting the rescaling of Eq.~\eqref{eq:gauge_eom_rescaled}, the physical gauge field amplitude $A_\pm$ is related to the rescaled field $u_\pm$ through the exponential of the integrated drag term. In radiation domination, the scale factor behaves as $a \propto x_\phi^{1/2}$ in dimensionless time, so that the integrated drag term evaluates to:
\begin{equation}
    \int^{x_\phi} \Gamma\left(\tilde{x}\right) \, d\tilde{x} = \frac{1}{2} \ln x_\phi + \frac{2 \beta}{M} \chi\left(x_\phi\right).
\end{equation}
The gauge field amplitude is
\begin{equation}
    A_\pm \left(x_{\phi,*}\right) = e^{-\frac{1}{2}\int^{x_{\phi,*}} \Gamma\,d\tilde{x}}\; u_\pm = x_{\phi,*}^{-1/4}\; e^{-\frac{\beta}{M}\chi_*}\; u_\pm\,,
\end{equation}
where $u_\pm$ is fixed by Bunch-Davies initial conditions and incorporates the full tachyonic growth history. The magnetic field carries an explicit factor of $I$ through the power spectrum per helicity mode,
\begin{equation}
    P_{B,\pm}(k) = \frac{I\, k^5 |A_\pm|^2}{4\pi^2 a^4}\,.
\end{equation}
The power spectrum is sharply peaked within the instability band, so the energy density
\begin{equation}
    \rho_{B,\pm} = \int \frac{dk}{k} P_{B,\pm}\left(k\right),
\end{equation}
is dominated by the peak mode $k_*$, so that $\rho_{B,\pm} \sim P_{B,\pm}$. Substituting $I = e^{2\beta\chi_*/M}$ and the expression for $A_\pm$, the dilaton prefactors cancel exactly
\begin{equation}
    B_{0,\pm}^2 
    \sim I\, k_*^5\, |A_\pm|^2.
\end{equation}
Our numerical results in Fig.~\ref{fig:MC_search_for_chi0} show that the magnetic field amplitude today is exponentially sensitive to the initial configuration of the dilaton field, with rapid amplification for initial values greater than $\chi_0 \sim 10^{-1}M$. The exponential sensitivity of the magnetic field amplitude today to the dilaton field arises from the dependence of the tachyonic growth exponent on $\beta \chi_*/M$ in the rescaled gauge field $u_\pm$.

%%%%%%%%%%%%%%%%%%%%%%%%%%%%%%%%%%%%%%%
\section{Dark Magnetic Field}
\label{sec:dark_magnetic_fields}
%%%%%%%%%%%%%%%%%%%%%%%%%%%%%%%%%%%%%%%
\par
In this work, the gauge field amplified should be understood as a dark $U(1)_D$ rather than the visible electromagnetic field~\cite{Choi:2018dqr}. This distinction is important because plasma conductivity during radiation domination strongly damps the visible-sector gauge-field growth~\cite{PhysRevD.56.5254}. A weakly coupled dark photon can undergo the axion-dilaton instability without feeling this obstruction directly, thus the observed intergalactic magnetic-field bounds are addressed only after kinetic mixing transfers a fraction of the generated dark magnetic field into the visible sector~\cite{PhysRevD.98.043501}.

\par
Due to the multi-frequency nature of the dispersion relation, the instability structure becomes analytically intractable as the scalar fields oscillate independent of each other, requiring a numerical treatment to explore the full parameter space of our model. For simplicity, we assume the scalar fields are oscillating coherently at early times with time-dependent amplitudes that decay with cosmic expansion. 

\par
We begin our simulation during radiation domination at a redshift of $z\sim10^5$, so that both scalar fields are coherently oscillating at initialization. The system is then evolved to matter-radiation equality. We choose an axion mass of $m_\phi =10^{-24} \, \mathrm{eV}$ with decay constant $f_\phi = 10^{26} \, \mathrm{eV}$ and coupling constant $\lambda=1$. We chose this fiducial axion mass value to maximize the hierarchy $H/m_\phi \ll 1$ during radiation domination, ensuring the dilaton friction of Eq.~\eqref{eq:GammaApprox} is well satisfied. However, we note the mechanism remains active across a much broader range of axion masses, as is demonstrated in Sec.~\ref{sec:axion_mass}. The initial misalignment angle is chosen to be $\theta_\phi \left(x_{\mathrm{init}}\right) = 40$, with initial velocity $\theta_\phi' \left(x_{\mathrm{init}}\right) = 0$. The axion angle is periodic, modulo $2\pi$~\cite{banerjee25, DiLuzio:2021pxd}. For the dilaton sector $m_\chi = r m_\phi$ with $r=1$ chosen for simplicity so that $m_\chi = m_\phi$. The coupling constants are chosen to be $\beta=-1$ and $M=M_\mathrm{Pl} = 1.22\times10^{28} \, \mathrm{eV}$ so that the dilaton coupling remains Planck-suppressed for values $\chi_0 /M \lesssim 1$. The dilaton field is initialized as $\chi_0 \approx 3.3\times 10^{27}\, \mathrm{eV}$ with initial velocity $\chi_0' =0$.

\par
\begin{figure}
    \centering
    \includegraphics[width=0.48\textwidth]{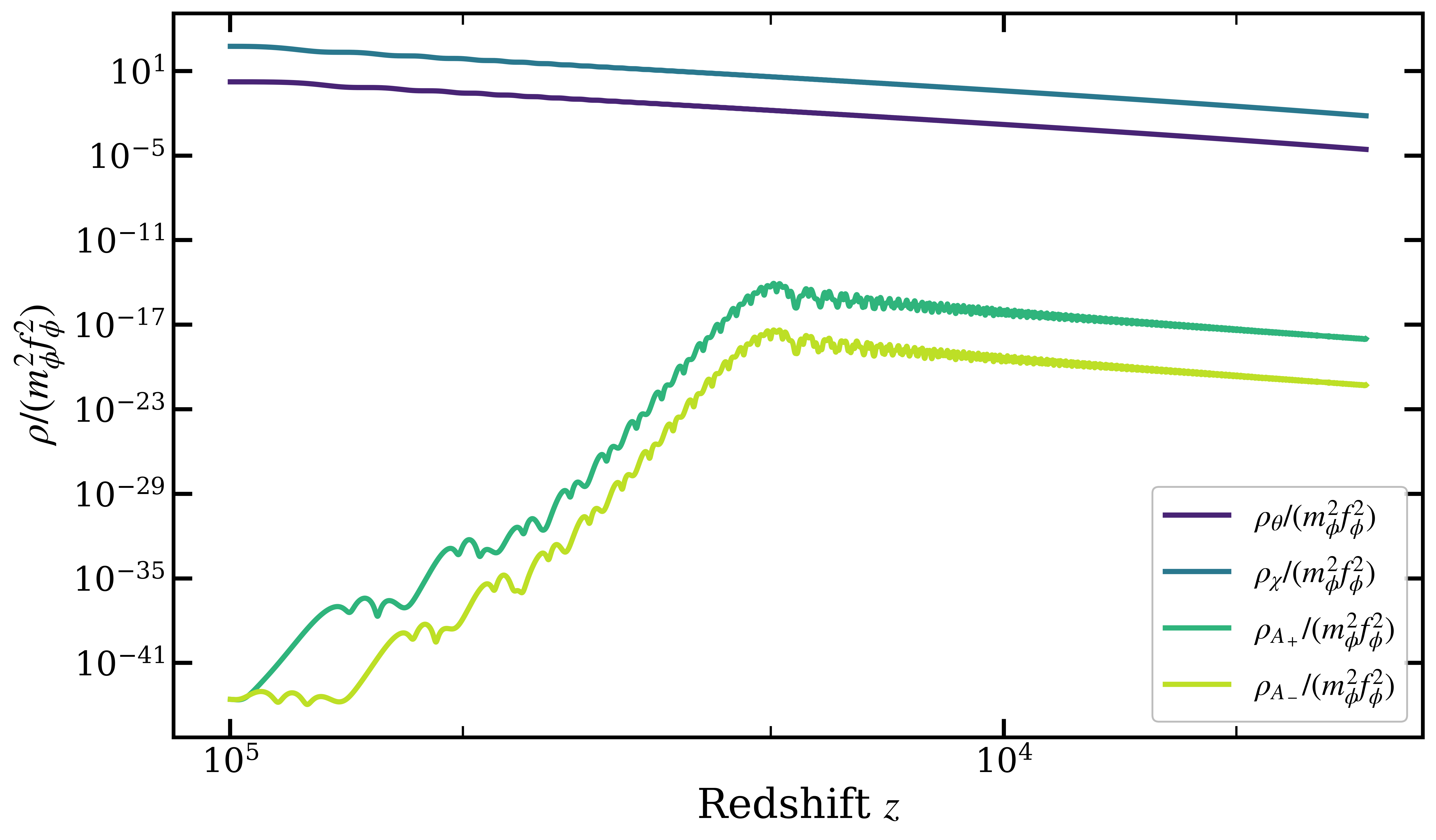}
    \includegraphics[width=0.48\textwidth]{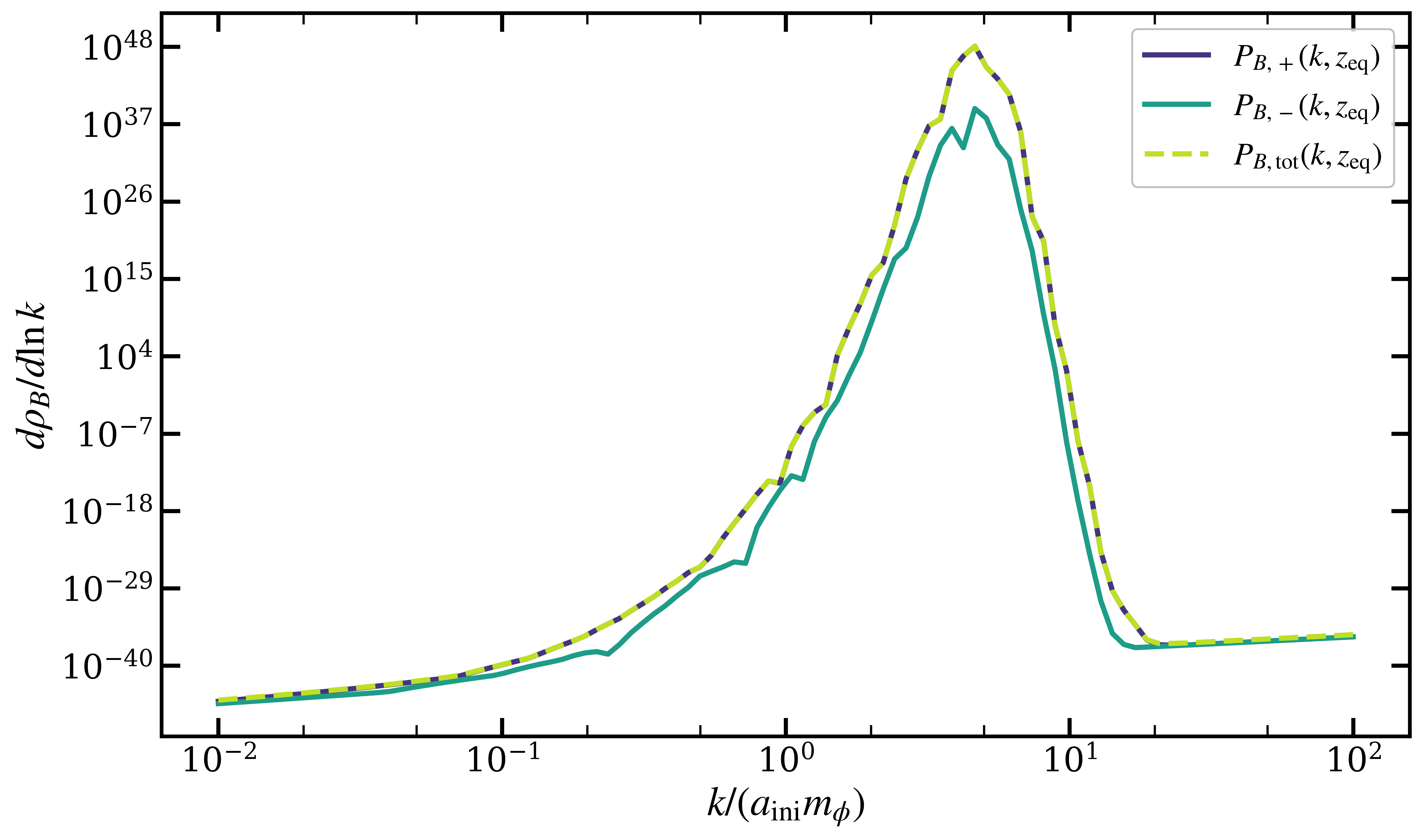}
    \caption{Top: Energy densities of the two scalar fields and both helicity modes of the dark field as a function of redshift for our benchmark parameters with $\beta=-1$. Bottom: The magnetic field power spectrum as a function of physical wavenumber $k_\mathrm{phys}$.}
    \label{fig:axi_dilaton_energy_density}
\end{figure}

The resulting evolution is shown in Fig.~\ref{fig:axi_dilaton_energy_density}. 
The top panel shows the energy densities of the two scalar fields and both helicity modes of the gauge field as a function of redshift. Both scalar fields decay as $\rho \propto a^{-3}$, which is consistent with the behavior of pressureless dark matter. The energy density of the gauge fields experiences a brief period of tachyonic amplification before peaking and decaying as radiation $\rho \propto a^{-4}$, while remaining subdominant to the background fields at all times. Both helicity modes $A_+$ and $A_-$ are simultaneously amplified, rather than a single preferred helicity mode, as predicted by standard axion models~\cite{Caprini:2014mja, garretson92}. Additionally, the periodic amplification of each helicity mode occurs out of phase. These features are a direct consequence of the dilatonic pump $C\left(x_\phi\right)$ discussed in Sec.~\ref{sec:the_model}, and confirm the analytic prediction that the $C > 0$ regime drives the amplification of both helicity modes within the finite band $0 < \kappa < \kappa_+$. 

\par
The bottom panel of Fig.~\ref{fig:axi_dilaton_energy_density} shows the resulting magnetic field power spectrum evaluated at the physical wavenumber $k/\left(a_{\mathrm{ini}} m_\phi\right)$. Here, $a_{\mathrm{ini}} \sim 10^{-5}$ is the value of the scale factor at our initial redshift. The figure shows a peak wavenumber of $k/\left(a_{\mathrm{ini}} m_\phi\right) \sim 4$, slightly shifted relative to the axion-driven scenario, which predicts a peak near $k/\left(a m_\phi\right) \sim 1$. We can compute the physical wavenumber today by, 
\begin{equation}
    \frac{k}{a_0} \sim 4 m_\phi \frac{a_{\mathrm{ini}}}{a_0} \sim 4 \times 10^{-29} \, \mathrm{eV}
\end{equation}
which corresponds to a physical length scale today of $\lambda_0 \sim 1 \, \mathrm{Mpc}$, consistent with current observational constraints. The two non-zero magnetic field power spectra are nearly coincident but not identical, showing the moderate chirality of the generated magnetic field. The amount of helicity can be quantified by the helicity fraction, 
\begin{equation}
    \varepsilon_H \equiv \frac{P_H}{P_B}
\end{equation}
where $P_B \propto \abs{B_{0,+}}^2 + \abs{B_{0,-}}^2$ is the magnetic field power spectrum and $P_H \propto \abs{B_{0,+}}^2 - \abs{B_{0,-}}^2$ is the helicity power spectrum, and $B_{0,\pm}$ is the amplitude of each mode of the helical magnetic field. Interestingly, the degree of chirality of the resulting magnetic field is sensitive to the relative amplification of each helicity mode, which itself depends on the parameter space of the model. In regions where each helicity mode is amplified with equal power, the net helicity fraction $\varepsilon_H \approx 0$, leaving behind a seemingly non-helical field despite the underlying helical amplification mechanism. A dedicated study of the helicity structure across the full parameter space is left for future work. 

\par
\begin{figure}
    \centering
    \includegraphics[width=0.48\textwidth]{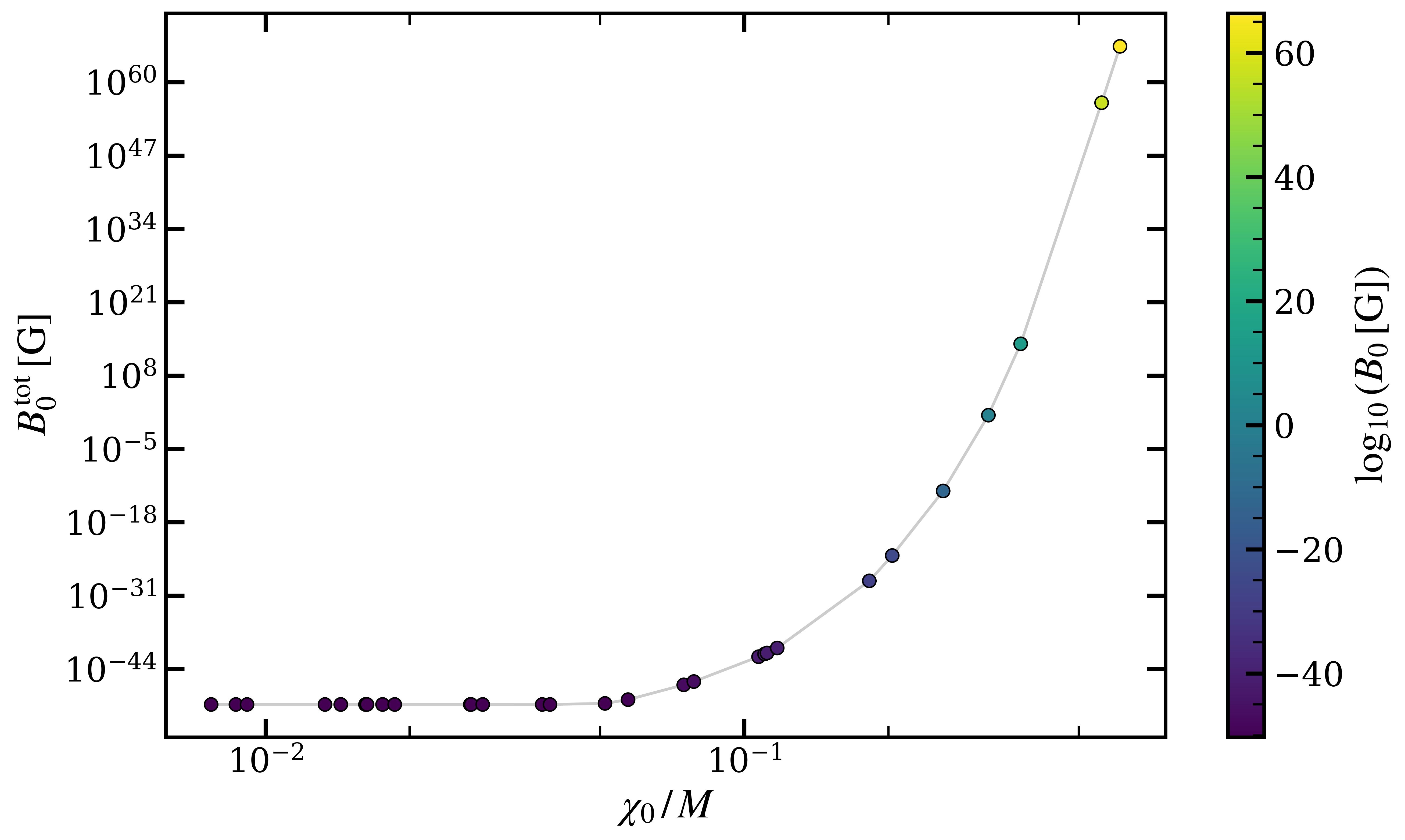}
    \caption{Parameter search over the initial dilaton value $\chi_0/M \in [0.007M_\mathrm{Pl},0.7M_\mathrm{Pl}]$ with $\beta=-1$. The dark magnetic field amplitude today grows rapidly as the initial value of the dilaton approaches the Planck mass.}
    \label{fig:MC_search_for_chi0}
\end{figure}

The exponential sensitivity of the magnetic field amplitude today to the value of the dilaton field at generation is confirmed numerically by the parameter search shown in Fig.~\ref{fig:MC_search_for_chi0}. We sample $N=30$ initial configurations for the dilaton field $\chi_0/M \in \left[0.007, 0.7\right]$ with coupling constant $\beta=-1$ and fixed initial misalignment angle value $\theta_0=40$, resulting in an observed magnetic field amplitude that grows rapidly as $\chi_0 \to M$, consistent with the predicted exponential scaling. The amplitude remains negligibly small for most of the sampled range before rising sharply near $\chi_0 \sim M$, confirming that the predicted enhancement of the magnetic field is an effect controlled primarily by the value of the dilaton rather than the tuning of the full parameter space. 
\begin{figure}
    \centering
    \includegraphics[width=0.48\textwidth]{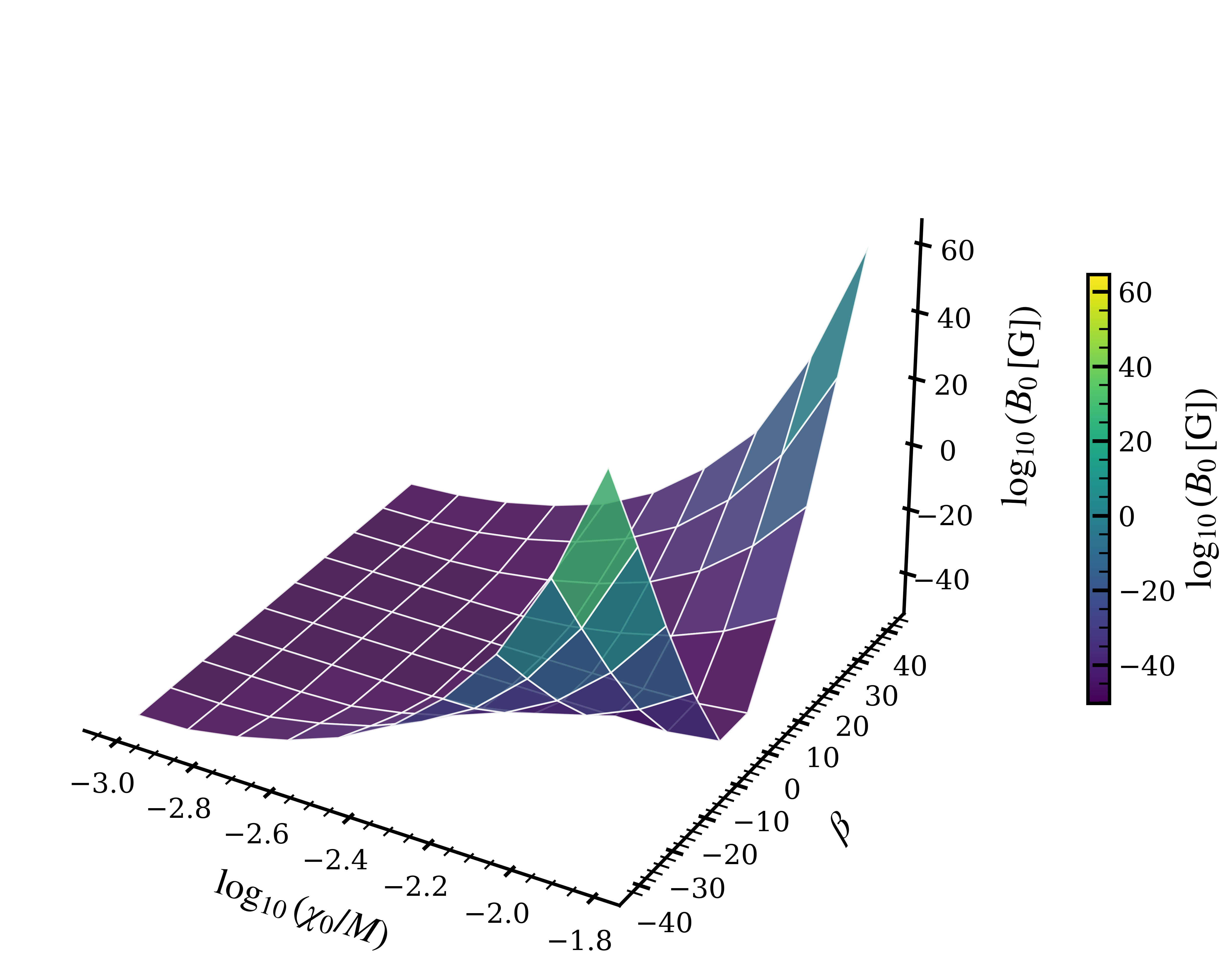}
    \caption{2D parameter search over $\chi_0/M$ and $\beta$, showing that the amplification is nearly symmetric about $\beta=0$ and persists across a wide range of coupling values.}
    \label{fig:2D_chi0_beta_parameter_search}
\end{figure}
\begin{figure}
    \centering
    \includegraphics[width=0.48\textwidth]{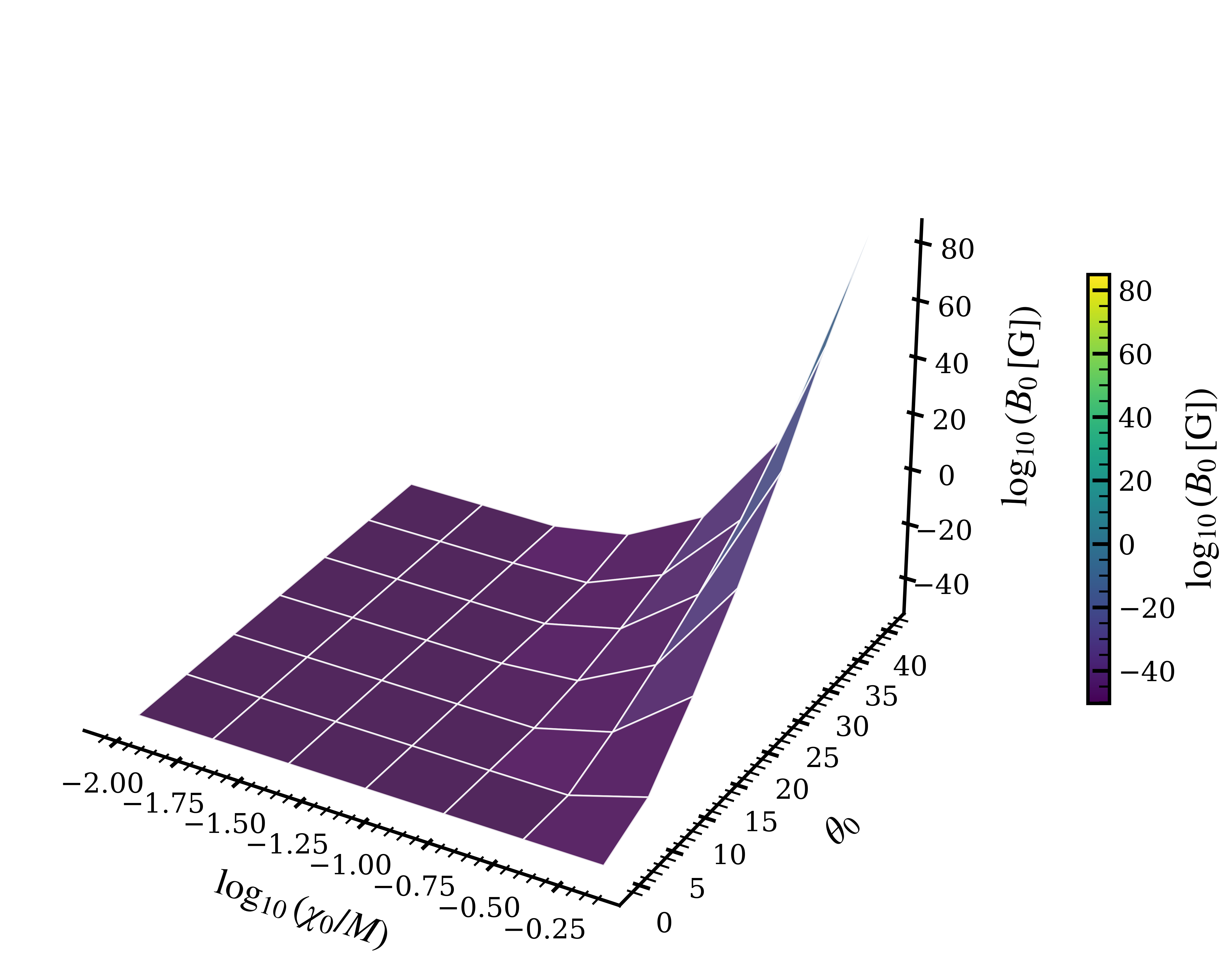}
    \caption{2D parameter search over the initial axion and dilaton field values, demonstrating that substantial amplification in our model requires contributions from both fields, with the dilaton playing the dominant role.}
    \label{fig:2D_chi0_theta0_parameter_search}
\end{figure}Fig.~\ref{fig:2D_chi0_beta_parameter_search} shows the two-dimensional parameter search over initial dilaton field configurations, $\chi_0/M$ and coupling $\beta$, confirming that the amplification is symmetric about $\beta=0$ and persists across a wide range of coupling values. Fig.~\ref{fig:2D_chi0_theta0_parameter_search} shows the joint dependence on the initial values the axion and dilaton fields, demonstrating that substantial amplification requires contributions from both the axion and dilaton sectors. 

\par
Although our model is valid for generic $U(1)$ field generation and amplification, the conductivity of plasma during radiation domination constrains any visible $U(1)$ sector~\cite{PhysRevD.56.5254}. 
\begin{equation}
\label{eq:conductivity}
    \sigma_\mathrm{phys}\sim\frac{T}{\alpha \ln(1/\alpha)}
\end{equation}
Accounting for the plasma, the gauge equations of motion gain an additional damping term, modifying the friction as $\tilde{\Gamma}= H/m_{\phi} + I'/I+\sigma_\mathrm{phys}/m_{\phi}$. After numerical evaluation, we find $\abs{I'/I}\ll\sigma_\mathrm{phys}/m_{\phi}$, showing that the conductivity of the plasma dominates over the dilatonic friction and washes out any amplification in the visible electromagnetic sector.

\par
Both the axion and dilaton fields behave as pressureless dark matter at late times. Their combined energy density at the end of radiation domination contributes to the total dark matter budget and cannot exceed $\rho_{\rm DM}=0.4\ \mathrm{eV}^{4}$~\cite{Planck:2018vyg}, constraining the initial conditions $\chi_0$, $\theta_0$ and the dilaton coupling $\beta$. Since the tachyonic growth exponent depends on the product $\beta\chi_0/M$ (cf.\ Eq.~(17)), the amplification can be held fixed while reducing the scalar energy density by lowering $\chi_0$ and compensating with a larger $|\beta|$. Imposing this constraint, a representative benchmark is found at $\beta=-25$, $\chi_0\approx 0.01M$, and $\theta_{0}=40$. The dark matter energy densities at the end of radiation domination, $z=3400$, satisfy
\begin{equation}
    \rho_{DM}=\rho_{\phi}+\rho_{\chi} =0.36\, \mathrm{eV}^{4}
\end{equation}
with $\rho_{\phi} = 0.26\, \mathrm{eV}^4$, $\rho_{\chi} = 0.10\, \mathrm{eV}^4$. The resulting value is below the observational threshold, at this benchmark the mechanism produces a dark magnetic field with amplitude $B_{0,\text{today}}=0.9 \, \mathrm{nG}$ with length scale $\lambda \sim 1 \, \mathrm{Mpc}$, showing that astrophysically relevant amplitudes can arise without fine-tuning either scalar sector. However, as emphasized by Eq.~\eqref{eq:conductivity}, the conductivity of the plasma at late times is large enough to dominate the Hubble friction in addition to the dilatonic damping (i.e. $\sigma_{\mathrm{phys}} \gg H, \Gamma$), ultimately placing any visible $U(1)$ gauge field in a strongly overdamped regime. This is not a limitation specific to the present model, but rather a generic obstruction for \emph{all} late-time magnetogenesis mechanisms that operate directly on the Standard Model photon~\cite{PhysRevD.56.5254}. Placing the gauge sector in a dark $U(1)_D$ removes this obstruction entirely, allowing the tachyonic instability to proceed unimpeded. 

\par 
The dark magnetic field can subsequently seed the observed intergalactic magnetic field through kinetic mixing with the Standard Model photon~\cite{Holdom:1986eq, 2026arXiv260522174N},
\begin{equation}\label{eq:kinetic_mixing}
    \mathcal{L} \supset -\frac{\epsilon}{2}\, F^{D}_{\mu\nu}\, F^{\mu\nu}_{\mathrm{EM}}\,,
\end{equation}
inducing a visible-sector magnetic field of amplitude
\begin{equation}\label{eq:Bvis}
    B_{\mathrm{vis}} \sim \epsilon\, B_{\mathrm{dark}}\,.
\end{equation}
Since kinetic mixing is linear at the level of the gauge fields, the power spectrum of the visible magnetic field retains the same coherence length scale and spectral shape as the dark field. We note that the recent work of Nandi and Choudhury~\cite{2026arXiv260522174N} also investigates how a dark-sector gauge field can participate in primordial magnetogenesis. Therein, the magnetic field is generated during inflation through a transient, time-dependent kinetic mixing between the photon and a dark photon, with the visible magnetic field sourced by the amplified dark-photon mode. Our mechanism is complementary, as the axion and dilaton control the overall tachyonic growth as well as the helicity structure, and kinetic mixing with the Standard Model photon acts as the transfer channel between the preexisting dark magnetic field and the visible sector.

\par
For ultralight dark photons there are no significant experimental constraints on the kinetic mixing parameter. CMB spectral distortion bounds, stellar cooling limits, and direct detection searches permit $\epsilon$ as large as $\mathcal{O}(10^{-1})$~\cite{PhysRevD.104.095029,Witte:2020rvb,An:2013yfc,Arias:2012az}. The present result of $B_{\mathrm{dark}} \simeq 0.9\,\mathrm{nG}$ provides substantial headroom for the transfer to the visible sector. Our result is consistent with current observational constraints, and a moderate mixing parameter $\epsilon \sim 10^{-1}$ would produce $B_{\mathrm{vis}} \sim 10^{-10}\,\mathrm{G}$, which is near the upper bound set by the CMB. The viable parameter space therefore spans approximately \emph{eight orders of magnitude} of the kinetic mixing parameter, and even larger by changing the initial condition, in stark contrast to single-field visible sector mechanisms, which typically require fine-tuned initial conditions or coupling constants to land within the observational window.

\par
We note that this broad viability is a direct consequence of the exponential amplification provided by the dilaton sector. Because the dark magnetic field already saturates the observational upper bound, the kinetic mixing parameter $\epsilon$ need only be large enough to exceed the lower bound, rather than being constrained to a narrow range. This separation of the amplification mechanism by the axi-dilaton from the sector-transfer mechanism through the kinetic mixing renders the model parametrically robust against uncertainties in either sector individually.

%%%%%%%%%%%%%%%%%%%%%%%%%%%%%%%%%%%%%%%
\section{Axion Mass Constraints}
\label{sec:axion_mass}
%%%%%%%%%%%%%%%%%%%%%%%%%%%%%%%%%%%%%%%
\begin{figure}
    \centering
    \includegraphics[width=0.48\textwidth]{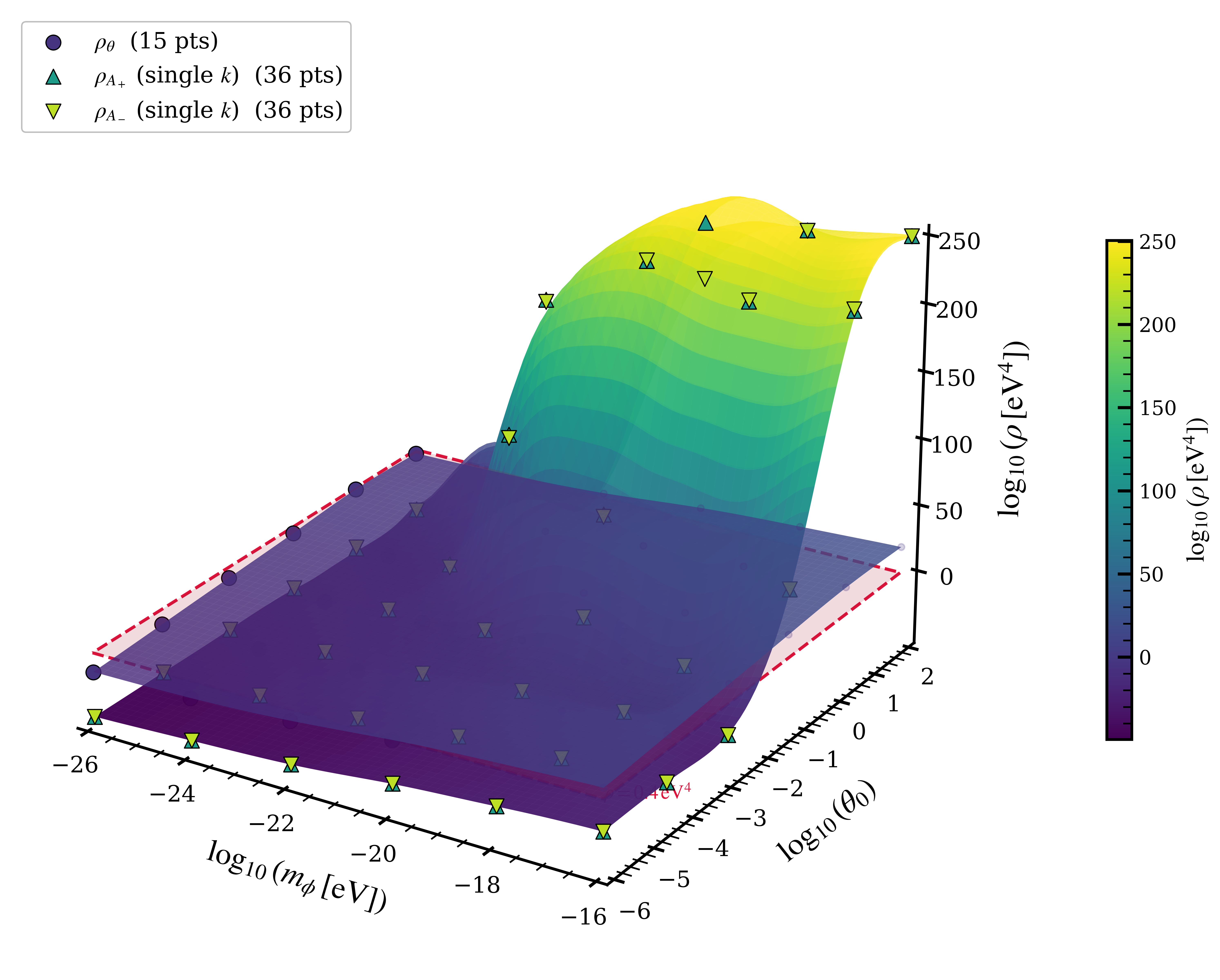}
    \caption{3D parameter scan of the dark gauge field energy density at matter-radiation equality as a function of the axion mass $m_\phi$ and the misalignment angle $\theta_0$. The red area shows the dark matter energy density constraint, $\rho \leq 0.4 \mathrm{eV}^4$.}
    \label{fig:axion_mass_figs}
\end{figure}

\begin{figure}
    \centering
    \includegraphics[width=0.48\textwidth]{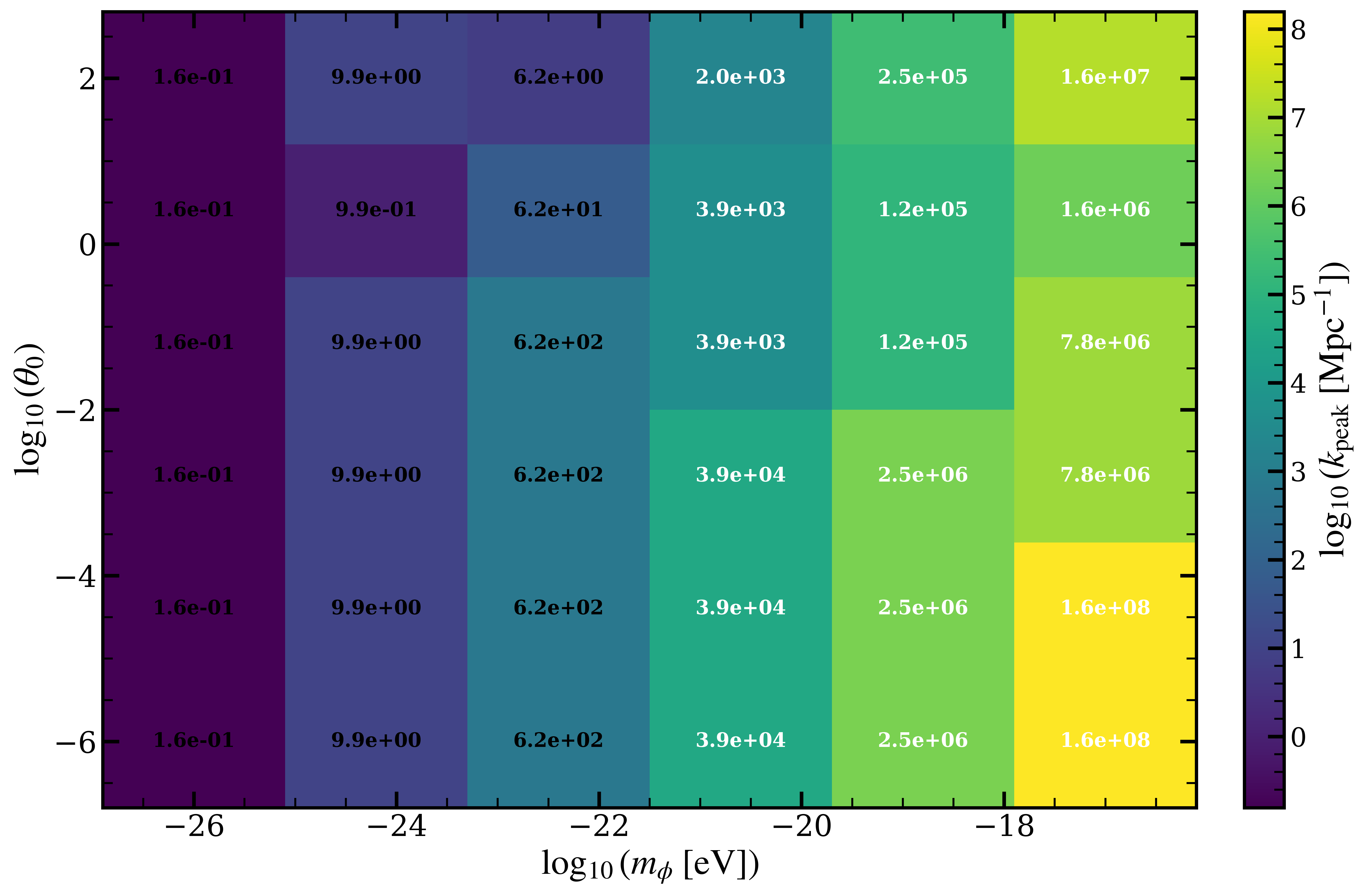}
    \caption{Parameter search over the initial misalignment angle $\theta_0$, the axion mass $m_\phi$, and the peak wavenumber $k_\mathrm{peak}$. Values of $k_\mathrm{peak}$ greater than $10^3$ are outside the scope of our model.}
    \label{fig:kpeak_shift}
\end{figure}

\begin{figure*}
    \centering
    \includegraphics[width=\textwidth]{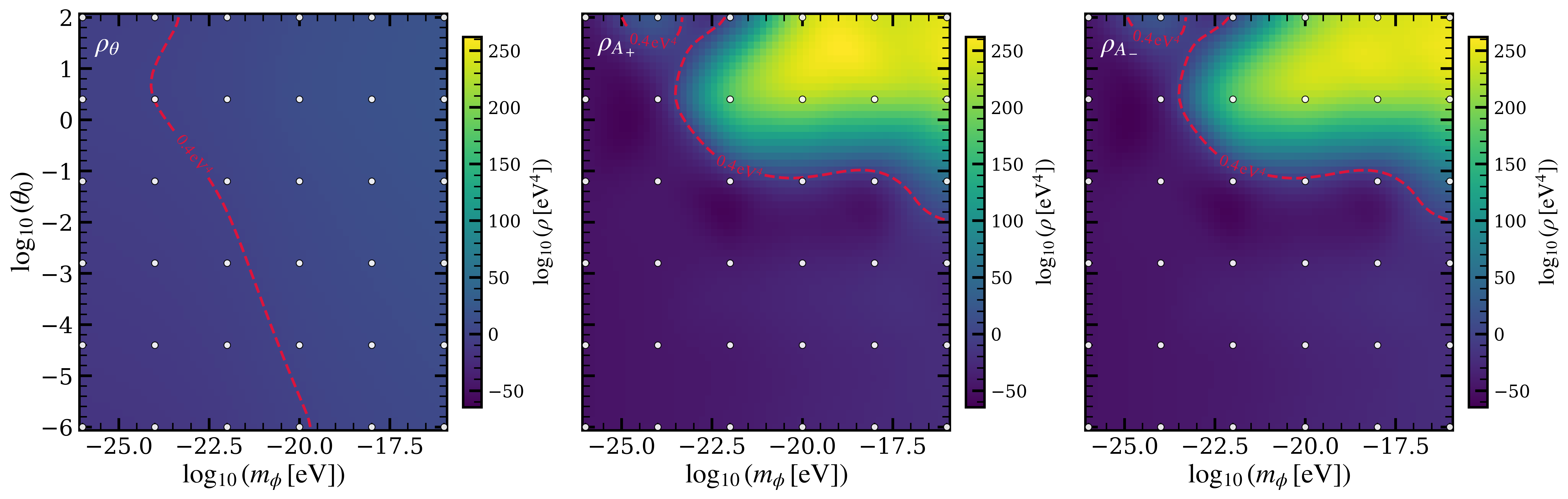}
    \caption{2D slice of the parameter scan of Fig.~\ref{fig:axion_mass_figs} showing the energy densities of the axion field $\rho_\theta$ (left), the $A_+$ helicity mode $\rho_{A_+}$ (center), and the $A_-$ helicity mode $\rho_{A_-}$ (right). The energy density budget constraint, $\rho \leq 0.4 \mathrm{eV}^4$ is shown as the red dashed line.}
    \label{fig:axion_mass_2d}
\end{figure*}

\par
The model presented in this work generates $\mathrm{Mpc}$-scale dark magnetic fields within the observationally relevant range, as shown in the previous section, section~\ref{sec:dark_magnetic_fields}. We now examine how this result depends on the axion mass, since it determines the physical scale set by the tachyonic instability. We fix the axion mass $m_\phi = 10^{-24} \,\mathrm{eV}$ for numerical evaluations. The parameter search discussed in this section then spans the range $m_\phi \in \left[10^{-16}, 10^{-26}\right]\, \mathrm{eV}$, covering the fuzzy dark matter and ultralight axion regimes~\cite{Hu:2000ke, PhysRevD.95.043541, PhysRevD.81.123530, Zelmer:2025gln}, and overlapping with the masses considered in recent late-time ultralight dark matter magnetogenesis models~\cite{Brandenberger:2025gks, Kamali:2026tgq}. The approximation in Eq.~\eqref{eq:GammaApprox} holds only when the dilaton friction dominates over the Hubble drag, and thus requires the condition $H/m_\phi \ll 1$ to be satisfied~\cite{garretson92}. For our benchmark value $m_\phi = 10^{-24} \mathrm{eV}$, we have $H/m_\phi \approx 10^{-9}$ at initialization. The evolution of the energy density shown in Fig.~\ref{fig:axi_dilaton_energy_density} confirms that the scalar fields decay as $\rho \approx a^{-3}$ throughout the simulation~\cite{banerjee25, Hu:2000ke, Hui:2021tkt}, consistent with coherent oscillation, and the backreaction analysis of App.~\ref{app:backreaction} confirms that the gauge field production does not disrupt this behavior.

\par
Figs.~\ref{fig:axion_mass_figs} and~\ref{fig:axion_mass_2d} show the result of a parameter scan over the axion mass and misalignment angle. Configurations with dark matter energy density beyond $\rho = 0.4 \, \mathrm{eV}^4$ are excluded by the observed dark matter density at recombination. Fig.~\ref{fig:axion_mass_figs} shows a 3D plot of the amplification landscape across the full mass range, demonstrating that substantial gauge field energy density is achieved over a broad region of the $\left(\theta_\phi, m_\phi\right)$ parameter space. Fig.~\ref{fig:axion_mass_2d} shows 2D slices of this landscape, illustrating the dependence on each parameter individually. These figures confirm that astrophysically relevant amplification is a generic feature of the parameter space, controlled by the combined action of the axion and dilaton sectors rather than by fine-tuning either the mass or initial angle individually. 

\par
Fig.~\ref{fig:kpeak_shift} shows the result of a parameter search over the initial misalignment angle, the axion mass, and the peak wavenumber $k_\mathrm{peak}$. This shows that the peak wavenumber depends more strongly on the axion mass than the initial misalignment angle, consistent with the analytic prediction that the tachyonic band is centered near $k \sim \mathcal{O}\left(1\right)$ in Eq.~\eqref{eq:dispersion_relation_quadratic_solution}, so that $k_\mathrm{peak}/a_\mathrm{initial} \sim m_\phi$. The scan covers the fuzzy dark matter and ultralight axion mass regimes, spanning approximately seven orders of magnitude from $m_\phi \in \left[10^{-18},10^{-25}\right] \, \mathrm{eV}$~\cite{Hu:2000ke, PhysRevD.95.043541, PhysRevD.81.123530, Zelmer:2025gln}. This clarifies where in the parameter space the mechanism is observationally distinctive. On these smaller coherence scales, astrophysical amplification mechanisms such as turbulent dynamos and structure-formation-driven magnetohydrodynamic evolution can produce comparable magnetic fields, making the primordial origin observationally degenerate~\cite{Donnert:2018wzc, Brandenburg:2004jv, Subramanian:2015lua, PhysRevD.70.123003, Kulsrud:1996km}. Consequently, a $\mathrm{kpc}$-scale measurement could not be unambiguously attributed to axi-dilaton magnetogenesis. The mechanism is therefore distinctive at the lower end of the mass range, matching the coherence length of intergalactic fields observed within voids, one of the primary observational motivations for invoking a primordial origin. The intergalactic ($\mathrm{Mpc}$) scale corresponds to the smaller mass $m \lesssim 10^{-22} \, \mathrm{eV}$, corresponding to the mass range of primary interest in this work. 

%%%%%%%%%%%%%%%%%%%%%%%%%%%%%%%%%%%%%%%
\section{Conclusion}
\label{sec:conclusion}
%%%%%%%%%%%%%%%%%%%%%%%%%%%%%%%%%%%%%%%
In this work, we present a late-time magnetogenesis mechanism driven by a coupled axion-dilaton system acting on a dark $U\left(1\right)$ gauge sector. Implemented during radiation domination, the model produces a dark magnetic field of astrophysically relevant amplitude, which can be subsequently transferred to the visible sector through kinetic mixing with the Standard Model photon, offering a broad viable parameter space for the mixing parameter $\epsilon$. The model combines a multiplicative dilaton kinetic coupling $I\left(\chi\right) = \exp\left(2\beta\chi/M\right)$, which serves as the primary driver of amplification, with a pseudoscalar coupling $\sim \theta_\phi F\tilde{F}$, which governs the helicity structure of the resulting field. 

\par
Numerical evolution from $z \sim 10^5$ to matter-radiation equality confirms these analytic predictions. Both helicity modes are amplified out of phase, and the magnetic field power spectrum peaks at a physical wavenumber corresponding to a coherence length scale $\lambda_0 \sim 1 \, \mathrm{Mpc}$ today. This is consistent with current observational constraints. A parameter search over the dilaton initial conditions (Fig.~\ref{fig:MC_search_for_chi0}) confirms the exponential sensitivity of the tachyonic growth history to the dilaton field value $\chi_0/M$, with the magnetic field amplitude rising sharply as $\chi_0 \to M$. The two-dimensional survey over $\chi_0$ and $\beta$ (Fig.~\ref{fig:2D_chi0_beta_parameter_search}) as well as $\chi_0$ and $\theta_0$ (Fig.~\ref{fig:2D_chi0_theta0_parameter_search}) further demonstrate that the amplification required to seed intergalactic magnetic fields is an extended feature of the parameter space of our model rather than a fine-tuned coincidence.

\par
The parameter search over the axion mass demonstrates that the viability of this mechanism is not limited to the fiducial value $m_\phi \sim 10^{-24} \, \mathrm{eV}$. Across the fuzzy dark matter and ultralight axion mass regimes, gauge field amplification remains effective over the $\left(\theta_0, m_\phi\right)$ parameter space, subject to the dark matter energy density constraint at matter-radiation equality. The mass dependence also clarifies the observational regime in which the model is most distinctive. Larger axion masses shift the spectrum to smaller coherence length scales where astrophysical dynamos and structure-formation processes can generate comparable fields. At the lower end of the mass range, $m_\phi \lesssim 10^{-22} \, \mathrm{eV}$, the mechanism predicts $\mathrm{Mpc}$-scale fields, consistent with magnetic fields in voids, which motivate a primordial origin.  

\par
Several directions remain for future investigation. First, a systematic study of the helicity structure across the full parameter space would determine whether the moderate chirality observed here is generic or confined to specific regions. Second, the backreaction of the amplified gauge field on the scalar background, which we have verified to be subdominant for the benchmark values considered, should be studied in greater generality. Third, a detailed treatment of the kinetic mixing transfer would sharpen the mapping between the dark magnetic field and the observable magnetic field. Fourth, the dilaton mass in our model could be chosen as dark energy such that it is consistent with cosmological observations at late times. Finally, the dark magnetic field itself may leave observable imprints through its effects on dark matter substructure, such as contributions to the effective number of relativistic species $\Delta N_{\mathrm{eff}}$~\cite{AtacamaCosmologyTelescope:2025nti, Planck:2018vyg}, unique lensing observables in the CMB~\cite{Shaw_2010, Saga:2023afb}, or signatures in the 21-cm absorption signal at high redshifts~\cite{schleicher08, sethi04, Worku:2026pmf}.

%%%%%%%%%%%%%%%%%%%%%%%%%%%%%%%%%%%%%%%
\section{Acknowledgments}
%%%%%%%%%%%%%%%%%%%%%%%%%%%%%%%%%%%%%%%
\par
We thank Professor Robert Brandenberger for comments and discussions on the early draft. 

%%%%%%%%%%%%%%%%%%%%%%%%%%%%%%%%%%%%%%%
\appendix
%%%%%%%%%%%%%%%%%%%%%%%%%%%%%%%%%%%%%%%
\section{Backreaction}
\label{app:backreaction}

\begin{figure}
    \centering
    \includegraphics[width=0.48\textwidth]{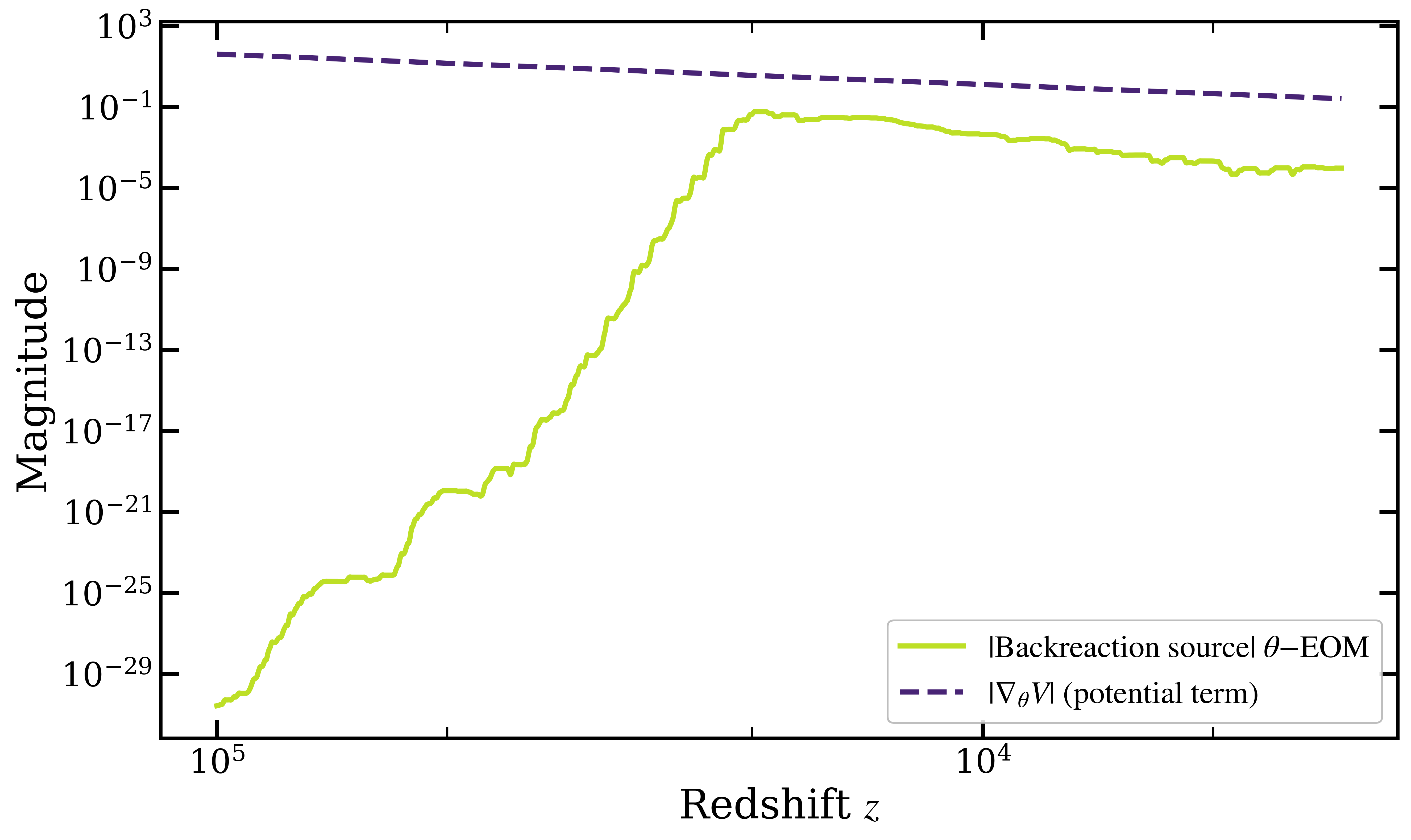}
    \includegraphics[width=0.48\textwidth]{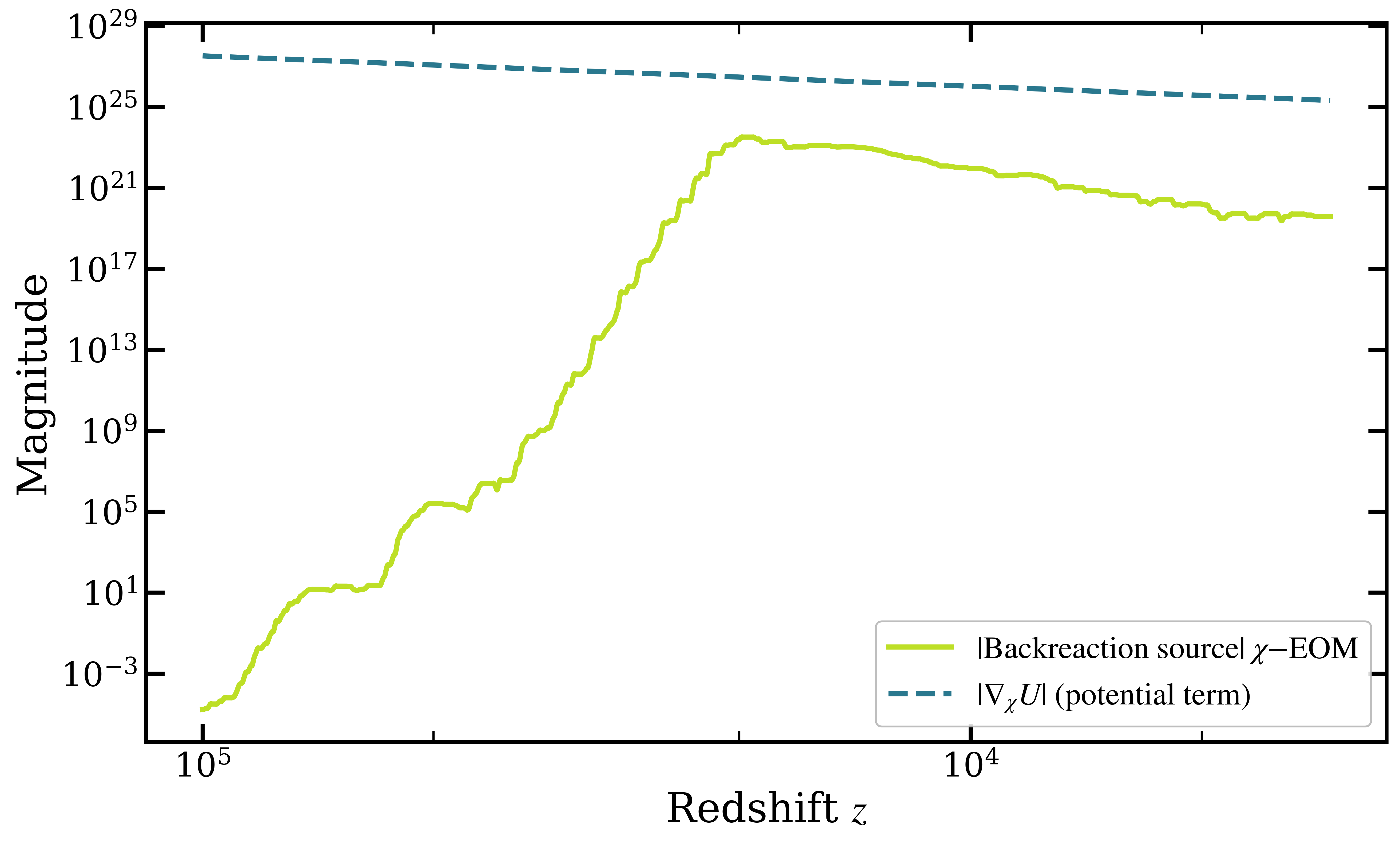}
    \caption{(Top) Backreaction source term for the axion compared to its potential in the axion equation of motion. (Bottom) Backreaction source term for the dilaton compared to its potential in the dilaton equation of motion.}
    \label{fig:axi_dilaton_backreaction}
\end{figure}

\par
In this work, we neglected the gauge-field backreaction on the scalar equations of motion. Here, we verify this approximation explicitly. Figure~\ref{fig:axi_dilaton_backreaction} compares the magnitude of the backreaction source terms, $\frac{\lambda}{4f_\phi^2 m_\phi^2} I(\chi) F_{\mu\nu}\tilde{F}^{\mu\nu}$ for the axion and $\frac{1}{m_\phi^2}\frac{\partial I}{\partial \chi}\left(\frac{1}{4}F^2 + \frac{\lambda}{4}\theta_\phi F\tilde{F}\right)$ for the dilaton, against the corresponding potential terms, cycle-averaged over the scalar oscillation period. In both sectors, the backreaction remains several orders of magnitude below the potential throughout the entire evolution, confirming that the coherent oscillation of the scalar backgrounds is not disrupted by gauge-field production.

\par
This hierarchy is independently corroborated by Fig.~\ref{fig:axi_dilaton_energy_density}, which shows that the energy density of the amplified gauge field stays subdominant to both scalar energy densities at all times. During the tachyonic amplification window the backreaction grows monotonically, reaching its maximum at the end of the amplification epoch. However, even at its peak the backreaction source remains negligible relative to the scalar potential. This indicates that its only effect is a slight temporal delay of the instability, rather than a significant suppression of the overall amplification.

\par
We also note that in unphysical regions of the parameter space where the resulting dark magnetic field amplitude $B_0$ becomes excessively large, the backreaction grows proportionally and can become comparable to the scalar potential, as expected from recent analytical studies of backreaction in ultralight-dark-matter magnetogenesis \cite{Brahma:2026fre}. In such regimes, the backreaction would act as a self-limiting mechanism, dynamically limiting the tachyonic growth and preventing the production of arbitrarily large dark magnetic fields. 

\par
We emphasize that the parameter searches are designed to map the full extent of the amplification landscape, particularly to demonstrate that astrophysically relevant magnetic field production is a generic feature of the parameter space rather than the result of fine-tuning. These surveys therefore do not incorporate the dynamical backreaction of the gauge field on the scalar backgrounds. In the regions of parameter space where the resulting $B_0$ reaches unphysically large values, the neglect of backreaction renders the quoted amplitudes unreliable. However, these extreme configurations lie well outside the astrophysically relevant regime and do not affect the conclusions drawn from the physically viable benchmarks discussed in Sec.~\ref{sec:dark_magnetic_fields}.

%%%%%%%%%%%%%%%%%%%%%%%%%%%%%%%%%%%%%%%
%\newpage
%\clearpage
\bibliographystyle{apsrev4-2}
\bibliography{bibliography}

\end{document}